\documentclass[letterpaper, 10 pt, conference]{ieeeconf}  

\usepackage{amsmath}
\usepackage{amssymb} 

\usepackage{tikz}
\usetikzlibrary{arrows,matrix,positioning,patterns}
\usetikzlibrary{calc}
\usepackage{pgfplots} 
\usetikzlibrary{plotmarks}
\usepackage[utf8]{inputenc}
\usepgfplotslibrary{groupplots}

\usepackage{colortbl}
\usepackage{url}
\usepackage{eurosym}
\usepackage{algorithm}

\usepackage{algorithm} 
\usepackage{algpseudocode}

\usepackage{bbold}
\DeclareSymbolFont{bbold}{U}{bbold}{m}{n}
\DeclareSymbolFontAlphabet{\mathbbold}{bbold}

\usepackage[labelformat=simple]{subcaption}
\DeclareCaptionLabelSeparator{periodspace}{.\quad}
\usepackage{array}\usepackage{calc} 

\IEEEoverridecommandlockouts                              
\title{\LARGE \bf
Stock Trading via Feedback Control:\\ Stochastic Model Predictive or Genetic?}

\author{Mogens Graf Plessen and Alberto Bemporad
\thanks{MGP and AB are with IMT School for Advanced Studies Lucca, Piazza S. Francesco 19, 55100 Lucca, Italy, {\tt\small \{mogens.plessen, alberto.bemporad\}@imtlucca.it}. Presented as a poster at the XVIII Workshop on Quantitative Finance 2017 (QFW2017).}%
}

\begin{document}

\maketitle
\thispagestyle{empty}
\pagestyle{empty}

\begin{abstract}
We seek a discussion about the most suitable feedback control structure for stock trading under the consideration of proportional transaction costs. Suitability refers to robustness and performance capability. Both are tested by considering different one-step ahead prediction qualities, including the ideal case, correct prediction of the direction of change in daily stock prices and the worst-case. Feedback control structures are partitioned into two general classes: stochastic model predictive control (SMPC) and genetic. For the former class three controllers are discussed, whereby it is distinguished between two Markowitz- and one dynamic hedging-inspired SMPC formulation. For the latter class five trading algorithms are disucssed, whereby it is distinguished between two different moving average (MA) based, two trading range (TR) based, and one strategy based on historical optimal (HistOpt) trajectories. This paper also gives a preliminary discussion about how modified dynamic hedging-inspired SMPC formulations may serve as alternatives to Markowitz portfolio optimization. The combinations of all of the eight controllers with five different one-step ahead prediction methods are backtested for daily trading of the 30 components of the German stock market index DAX for the time period between November 27, 2015 and November 25, 2016.     

\end{abstract}

\section{Introduction}

Within the context of performance-related asset trading, we distinuish between three general tasks: system identification (finding of cause and effect relations), scenario generation (future asset price predictions) and trade decision taking (control logic). This paper focuses on the third task. For low-level trading mechanics and feedback control thereof we refer to \cite{barmish2015stock}. For interesting recent  control theory-related research problems in finance associated with the control of order book dynamics, see \cite{barmish2016nasdaq}. This work is founded on \cite{markowitz1952portfolio} and \cite{bemporad2011stochastic}. The motivation for this paper is the intention to extend a stochastic model predictive control approach to multiple-asset portfolio optimization for profit- and risk-related objectives. However, the suitability of SMPC needs first to be compared to alternative control strategies. Such a comparison is provided below. The main contribution of this paper is thus analysis to find the most suitable general feedback control structure for stock trading out of  two general and large classes: stochastic model predictive control (SMPC) and genetic algorithms. In this paper, we here refer to a genetic algorithm as any customized control method of arbitrary structure whose parameters are optimized through simulation using real-world data.

We compare eight different stock trading algorithms that can be partitioned into the two aforementioned classes. For scenario generation, on which we rely all controllers, we assume five different one-step ahead price prediction methods. Their quality ranges from ideal (perfect price-ahead prediction) to totally off (wrong price rate sign-prediction at all sampling intervals). It is stressed that we explicitly do \textit{not} consider multi-asset \textit{portfolio optimization}, but instead concentrate on the trading of separate \textit{single assets} for a given period of time. The used real-world data is drawn from the 30 components of the German stock market index DAX.

The remainder of this paper is organized as follows. System dynamics are described in Section \ref{sec_transitionDynamics}. Section \ref{sec_SMPC} introduces two different stochastic model predictive stock trading frameworks.  Genetic stock trading algorithms are outlined in Section \ref{sec_geneticTrading}. Simulation experiments are reported in Section \ref{sec_expts} before concluding with Section \ref{sec_conclusion}. \\

\section{Transition Dynamics Modeling}\label{sec_transitionDynamics}

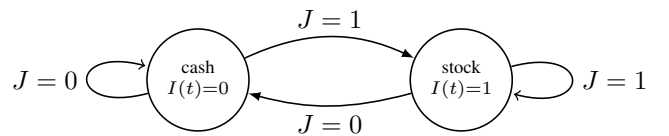
\begin{figure}[htbp]
\centering  
\begin {tikzpicture}[-latex ,auto ,node distance =2 cm and 3.5cm ,on grid ,
semithick ,
state/.style ={ circle ,top color =white , bottom color = white ,
draw, text=black, minimum width =1 cm}]
\node[state] (A) {$\text{cash}\atop I(t)=0$};
\node[state] (B) [right =of A] {$\text{stock}\atop I(t)=1$};
\path (A) edge [loop left] node[left] {$\tiny{J=0}$} (A);
\path (A) edge [bend left =25] node[above] {$\tiny{J=1}$} (B);
\path (B) edge [bend left =15] node[below =0 cm] {$\tiny{J=0}$} (A);
\path (B) edge [loop right] node[right] {$\tiny{J=1}$} (B);
\end{tikzpicture}
\caption{Visualization of the Markov decision process (MDP) when trading cash and a stock only.}
\label{fig:MDP}
\end{figure}

Let time index $t$ be associated with sampling time $T_s$ such that all time instances of interest can be described as $tT_s$, whereby, throughout this papers, we have $T_s=1$ day.  Let us define the system state by
\begin{equation}
Z(t)= \begin{bmatrix} I(t) & M(t) & N(t) & W(t) \end{bmatrix}^T,\label{eq_def_zt}
\end{equation}
with $I(t)\in\{0,1\}$ indicating a cash- or stock-investment, respectively, $M(t)\in\mathbb{R}$ the current cash position (measured in currency \EUR), $N(t)\in\mathbb{Z}_+$ the number of shares held, and $W(t)\in\mathbb{R}_+$ the current portfolio wealth. Thus, to analyze a suitable stock trading algorithm, we assume no fractional investments being possible, but either entirely cash- or stock-investment. Transition dynamics can then be modeled as a Markov decision process (MDP). Control variable $J(t)\in\{0,1\}$ decides upon investment positions according to Figure \ref{fig:MDP}. In general, we model transaction costs as non-convex with  a fixed charge for any nonzero trade (fixed transaction costs) and a linear term scaling with the quantity traded (proportional transaction costs). Thus, at time $t-1$, the purchase of $N(t-1)$ shares of an asset
 results in $M(t)=M(t-1) - N(t-1)s(t-1)\left(1+\epsilon_\text{buy}\right) - \beta_\text{buy}$, with $s(t)$ denoting asset (closing) price at time $t$. Similarly, for the selling of $N(t-1)$ assets we have $M(t)=M(t-1)+N(t-1)s(t-1)\left(1-\epsilon_\text{sell}\right)
-\beta_\text{sell}$. For the remainder of this paper we assume no fixed transaction costs, i.e., $\beta_\text{sell}=\beta_\text{buy}=0$. This simplification is done to directly adapt the convex problem formulations from \cite{markowitz1952portfolio} and for the dynamic hedging-inspired formulation proposed in Section \ref{sec_SMPC}. Fixed transaction costs, that render the problem non-convex, can be approached by iterative relaxation methods \cite{lobo2007portfolio} or hybrid system theory. For wealth dynamics, we have $W(t)\in\left\{W(t-1),M(t),\tilde{W}(t)\right\}$, whereby $\tilde{W}(t)=\tilde{M}(t) + \tilde{N}(t)s(t)$ with $\tilde{M}(t)$ denoting the optimizer and $\tilde{N}(t)$ the optimal objective function value of 
\begin{equation*}
\max_{M(t)\geq 0}\left\{N(t)\in\mathbb{Z}_+: N(t)=\frac{M(t-1)-\beta_\text{buy} - M(t)}{s(t-1)\left(1+\epsilon_\text{buy}\right)} \right\}.
\end{equation*} 
Thus, given $M(t-1)$, we find the largest possible positive integer number of assets we can purchase under consideration of transaction costs. The smallest possible cash residual is $\tilde{M}(t)=0$.

Within this paper, the focus is on  \textit{unconstrained} trading frequency of \textit{one} asset, i.e., trading is permitted on any two consecutive trading days, and confining cash and asset to be based on the \textit{same} currency. More general is the treatment of multiple assets, multiple foreign exchange rates (\textit{forex}), and various constraints such as a bound on the total number of admissable trades, a waiting period in between two consecutive trades or diversification constraints, which require a state-space extension but or not subject of this paper explicitly. 

To summarize, at every trading interval $t$ we conduct following algorithm:
\begin{enumerate}
\item Read current $s(t)$ to update $W(t)$ and thus $Z(t)$.
\item Decide on $J(t)\in\{0,1\}$.
\item Rebalance the portfolio according to $J(t)$.
\end{enumerate}
All of the following two sections are concerned about the decision on $J(t)\in\{0,1\}$ with fundamental objective profit maximization.

\section{Stochastic Model Predictive Stock Trading} \label{sec_SMPC}

\newlength\figureheight
\newlength\figurewidth 
\setlength\figureheight{5cm}
\setlength\figurewidth{8.75cm}
\begin{figure*}
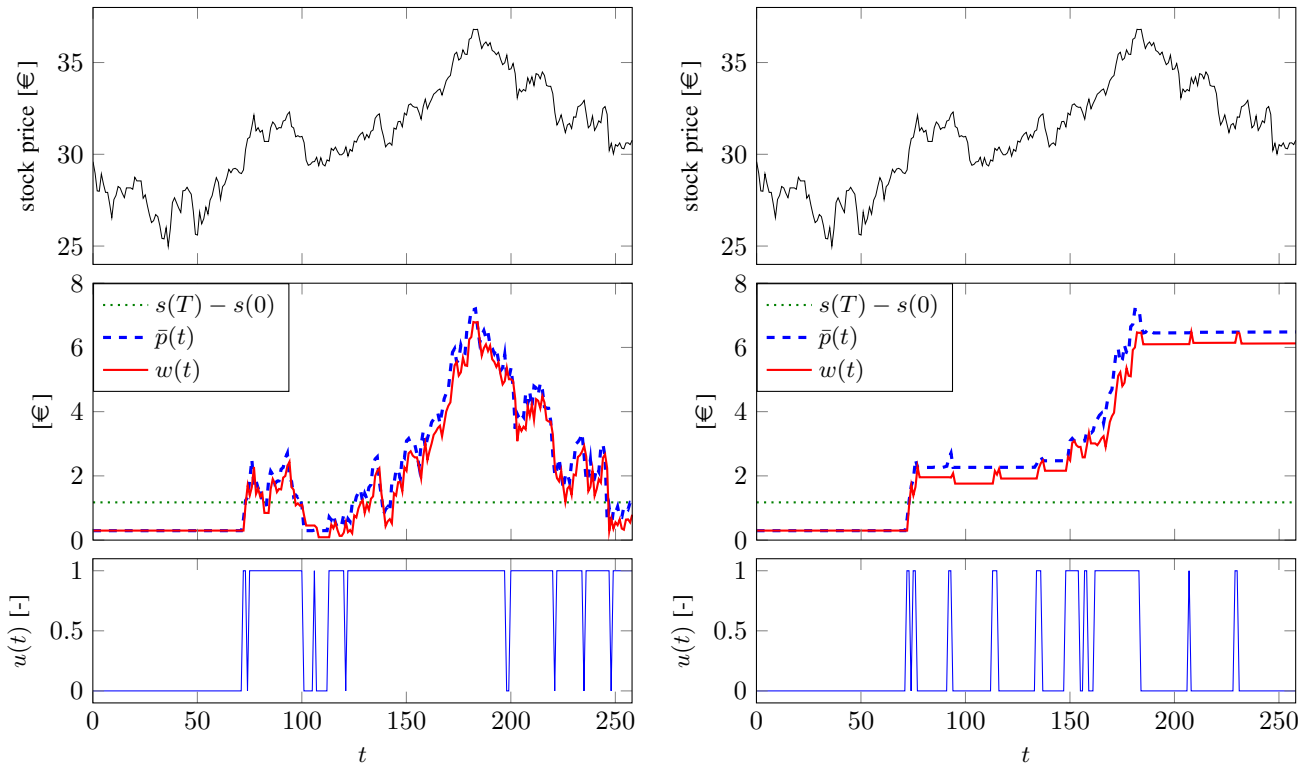

\centering
\input{fig_hedging.tex}
\input{fig_portf.tex}
\caption{Visualization of the dynamic hedging-inspired SMPC formulation for stock trading. A noncausal setting with perfect $s(t+1)$ knowledge is assumed for illustration. Parameters for scenario generation and perturation noise are $(M,\sigma_\text{pert})=(100,0.3)$. (Left) Dynamic hedging result. (Right) Result of \textsf{SMPC-DH}. In both cases, \eqref{eq_OP_LPMinMax} is solved. The difference is the generation of references $p^j(t+1),~j=1,\dots,M$, see Section \ref{subsec_SMPC_DH}. The average is denoted by $\bar{p}(t)=\frac{1}{M}\sum_{j=1}^M p^j(t)$. The underlying stock (top frame) is of Vonovia SE (November 27, 2015 until November 25, 2016).}
\label{fig:SMPC_DH}
\end{figure*}
\setlength\figureheight{7cm}
\setlength\figurewidth{8.75cm}
\begin{figure}
\centering
\input{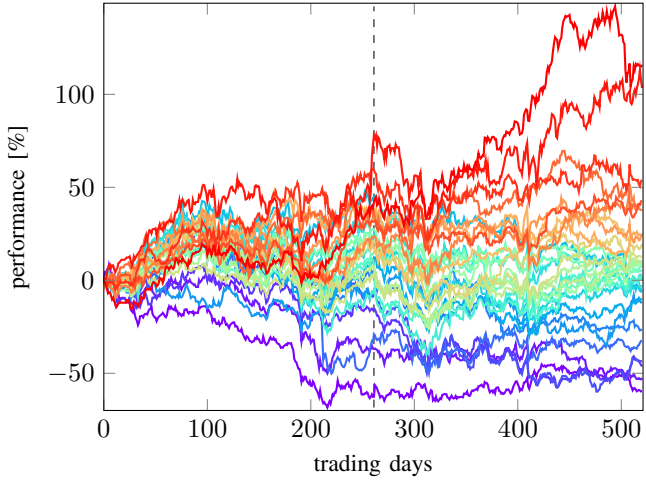}
\caption{Normalized two-year evolution of all 30 components held in the DAX between November 28, 2014, and November 25, 2016. The data is partitioned by the black-dashed vertical line into training and evaluation data, respectively. Thus, $t=0$ is initialized at trading day 261 (November 27, 2015).}
\label{fig:fig1}
\end{figure} 

Let us discuss a relation between portfolio optimization and dynamic hedging using stochastic model predictive control (SMPC). For a financial institution, hedging a derivative contract implies to dynamically rebalance a so-called \textit{replicating} portfolio of underlying assets at periodic intervals so that, at the expiration date of the contract, the value of the portfolio is as close as possible to the payoff value to pay to the customer. For the multiple asset replicating portfolio case, wealth dynamics $w(t)$ of the replicating portfolio can be defined as 
\begin{equation}
w(t+1) = (1+r)\left( w(t) - \sum_{i=1}^n h_i(t) \right) + \sum_{i=1}^n b_i(t) u_i(t)\label{eq:wealth-after-with-tc}
\end{equation} 
where $u_i(t)$ is the \textit{fractional} quantity of asset $i$ held at time $t$, $b_i(t)=s_i(t+1)-(1+r)s_i(t)$ is the excess return, i.e., how much the asset gains (or loses) with respect to the risk-free rate $r$ over interval $T_s$, and transaction costs $h_i(t)$ are assumed to be proportional to the traded quantity of stock, i.e., 
\begin{equation}
h_i(t)=\epsilon_i s_i(t) |u_i(t)-u_i(t-1) |,\label{eq:prop-costs1}
\end{equation}
with fixed quantity $\epsilon_i \geq0$ depending on commissions on trading asset $i$, $i=1,\dots,n$ (we assume no costs are applied on trasacting the risk-free asset). A standard option contract typically covers 100 shares. Thus, $u_i(t) = 1$ implies a portfolio such that at the end of the rebalancing interval 100 shares of underlying asset $i$ are held. For our setting, we here assume one asset only and drop subscripts correspondingly. Furthermore, we set $\epsilon=\epsilon_\text{buy}$ and $\epsilon_\text{sell}=\epsilon_\text{buy}$. For the formulation of convex optimization problems, we introduce a \textit{virtual} portfolio, constrain\footnote{The unconstrained case admitting $u(t)<0$ would imply the possibility of \textit{shortselling} stocks.} $u(t)\in\{0,1\}$, and then relate $I(t)=u(t)$.

\subsection{Two Markowitz-inspired SMPCs}

With regard of portfolio optimization, Markowitz \cite{markowitz1952portfolio} trades-off the mean (performance) and variance (risk) of the return. For our setting, this objective can be formulated as
\begin{align}
\max_{u(t)\in\{0,1\}} &\ \ \text{E}[w(t+1)] - \frac{\alpha}{2} \text{Var}[w(t+1)], \label{eq_OP_objfcn}
\end{align}
where $\alpha$ denotes the trade-off parameter. The decision of selecting $u(t)$ is largely dependent on $s(t+1)$, which is unknown at time $t$. Employing a SMPC approach, we therefore generate $M$ scenarios for possible future prices $s^j(t+1)$ with corresponding probabilities $\pi^j,~j=1,\dots,M$. Accordingly, we obtain $w^j(t+1)$, $\text{E}[w(t+1)]=\sum_{j=1}^M \pi^j w^j(t+1)$ and $\text{Var}[w(t+1)]=\text{E}[w^2(t+1)] - \text{E}^2[w(t+1)]$. We generate scenarios as 
\begin{equation}
s^j(t+1) = \hat{s}(t+1) + \sigma_\text{pert} \eta^j(t),~\eta^j(t)\sim\mathcal{N}(0,1),\label{eq_def_sjtp1}
\end{equation}
where $\hat{s}(t+1)$ denotes our mean estimate of $s(t+1)$ and $\sigma_\text{pert}\geq 0$ is a tuning parameter to add \textit{perturbation noise}. For final experiments we assume $M=100$. Thus, our first SMPC-based controller solves \eqref{eq_OP_objfcn} with scenario generation according \eqref{eq_def_sjtp1}. It is referred to as \textsf{SMPC-M100}. For the \textit{fractional} case relevant for dynamic option hedging (and specific for $\Delta$-hedging \cite{BS73}), the corresponding to \eqref{eq_OP_objfcn} can be cast into a quadratic program (QP) by the introduction of two slack variables. For our case, we just evaluate the objective function for both $u(t)\in\{0,1\}$ and therefore do not require a QP-solver. 

In a second setting we assume $M=1$ which implies $\text{Var}[w(t+1)]=0$. In order to still maintain a possible knob to trade-off performance and risk, we define
\begin{align}
\max_{u(t)\in\{0,1\}} &\ \ \text{E}[w(t+1)] - \frac{\beta}{2} (u(t)-u(t-1))^2 \sigma, \label{eq_OP_Eplus}
\end{align}
where $\beta$ is a tuning parameter and $\sigma$ was introduced to relate to data. Assuming the log-normal stock model, it is typically estimated as the \textit{maximum likelihood} (ML) from $\mathcal{T}+1$ past stock prices using $\{\ln(\frac{s(t-\mathcal{T}+1)}{s(t-\mathcal{T})}),\dots,\ln(\frac{s(t)}{s(t-1)}) \}$. The higher $\beta$, the less frequent $u(t)$ is varied. Again, we solve \eqref{eq_OP_Eplus} by evaluation for both $u(t)\in\{0,1\}$ instead of solving the more general QP. We refer to the resulting controller as \textsf{SMPC-E+}, eventhough, strictly it is not stochastic anymore since $M=1$.

\subsection{Dynamic hedging-inspired SMPC}\label{subsec_SMPC_DH}

For option hedging, the objective is to typically minimize the so-called hedging error $e(T)=w(T)-p(T)$, where $w(T)$ and $p(T)$ denote replicating portfolio wealth and option price at expiration date $T$, respectively. Guiding notion of \textit{dynamic} option hedging is to minimize the ``tracking error'' $e(t)=w(t)-p(t),~\forall t=0,\dots,T$ for all possible asset price realization. In order to minimize under transaction costs, three different stochastic measures of the predicted hedging error are discussed in \cite{bemporad2011stochastic}. We here focus on the \textsf{LP-MinMax} formulation, minimizing the maximal hedging error resulting from scenario generation, i.e.,
\begin{align}
\min_{u(t)\in\{0,1\}}\max_{j=1,\dots,M} &\ \ |w^j(t+1)-p^j(t+1) |. \label{eq_OP_LPMinMax}
\end{align}
Its benefit over the other two stochastic measures is independence from any trade-off parameter. Let us discuss how the framework \eqref{eq_OP_LPMinMax} can serve for stock trading. First, for hedging of a European call option, the analytical scheme to generatio option price scenarios is $p^j(t+1)=(1+r)^{-(T-(t+1))} \max(s^j(T)-K_s,0),~j=1,\dots,M$. The option strike price is denoted by $K_s$. For our stock trading objective, we modify these ``references''. For \eqref{eq_OP_LPMinMax}, we therefore propose  
\begin{equation}
p^j(t+1) = (1+r)^{-(T-(t+1))} \max(s^j(t+1)-s(0),l(t)),\label{eq_def_pjtp1_portf}
\end{equation}
with 
\begin{equation*}
l(t) = \begin{cases} \max(w(t),~0), & \text{if}~w(t)>l(t),\\
l(t-1),& \text{otherwise}, \end{cases}
\end{equation*}
and initialize $l(0)=0$. The corresponding dynamic hedging-inspired controller shall be referred to as \textsf{SMPC-DH}. Both the difference with respect to dynamic option hedging and the motivation for employing \eqref{eq_def_pjtp1_portf} for stock trading are visualized in Figure \ref{fig:SMPC_DH}. Using \eqref{eq_def_pjtp1_portf} in combination with \eqref{eq_OP_LPMinMax} can be interpreted as a \textit{trailing stop-loss} strategy. Finally, we remark that the other two stochastic measures (\textsf{QP-Var} and \textsf{LP-CVaR}) from \cite{bemporad2011stochastic} can be employed likewise using \eqref{eq_def_pjtp1_portf}.


\section{Genetic Stock Trading}\label{sec_geneticTrading}

We discuss five genetic stock trading methods. 

\subsection{Two moving average-based controllers}

Let us define the moving average (MA) of a stock price as $s_\text{MA}(t+1)= \frac{1}{p_\text{MA}} \sum_{\tau=0}^{p_\text{MA}-1} \tilde{s}(t+1-\tau)$, with $\tilde{s}(t+1)=\hat{s}(t+1)$ and $\tilde{s}(t+1-\tau)=s(t+1-\tau),~\forall \tau\geq 1$, and where $p_\text{MA}$ is the moving average length parameter. The first MA-based controller, referred to as \textsf{MA-Cross} in the following, triggers a buy-signal ($J(t)=1$) in case of a short-term MA coming from below crossing a long-term MA. Similarly, a sell-signal ($J(t)=0$) is generated in case of the short-term MA crossing the long-term MA from above. The two parameters defining MA-lengths shall be denoted by $p_\text{MA,s}$ and $p_\text{MA,l}$.

The second MA-based controller, below referred to as \textsf{MA-Sign}, takes as input parameters $p_\text{MA}$ and $T_\text{MA}$. It then computes $\Delta s_\text{MA}(t-\tau) = s_\text{MA}(t+1-\tau)- s_\text{MA}(t-\tau),~\forall \tau=0,1,\dots,T_\text{MA}-2$. A buy-signal is generated if $\text{sign}\left(\Delta s_\text{MA}(t-\tau)\right)>0,~\forall \tau=0,1,\dots,T_\text{MA}-2$, a sell-signal otherwise, and where $\text{sign}(\cdot)$ denotes the sign-operator. The guiding idea is to exploit price trends using constant-sign MA-slope rates for the past $T_\text{MA}-1$ intervals. We refer to this second MA-based controller as \textsf{MA-Sign}.

\subsection{Two trading range-based controllers}

Let us select a time window $[t-T_\text{win},t]$ and partition it such that 
\begin{equation}
T_\text{win}=K p_\text{TR} + \delta,\label{eq_def_Twin}
\end{equation}
where $p_\text{TR}\in\mathbb{Z}_{++}$ is a parameter, $K\in\mathbb{Z}_{++}$, and $\delta\geq 0$ a corresponding residual of time-instances. We  define interval-wise \textit{local maxima} by
\begin{equation}
s_\text{max}^{(k)} = \max_{\tau\in[t-T_\text{win}+(k-1)K,t-T_\text{win}+kK]} s(\tau), \label{eq_def_smax_k}
\end{equation}
and the corresponding time arguments by $t_\text{max}^{(k)},~\forall k=1,\dots,K$. Similarly we derive \textit{local minima} $s_\text{min}^{(k)}$ and $t_\text{min}^{(k)}$. Local minima and maxima are suitable to generate \textit{trading ranges} (TR). We refer to our first TR-based controller as \textsf{TR-Inside}. It triggers trading signals as follows:
\begin{equation*}
J(t) = \begin{cases} 0, & \text{if}~\frac{|\hat{s}(t+1)-\hat{y}_\text{max}(t+1)|}{\hat{y}_\text{max}(t+1)}<\epsilon_\text{TR},\\
1, & \text{if}~\frac{|\hat{s}(t+1)-\hat{y}_\text{min}(t+1)|}{\hat{y}_\text{min}(t+1)}<\epsilon_\text{TR},\\
J(t-1), & \text{otherwise},\end{cases}
\end{equation*} 
where $\hat{y}_\text{max}(t+1) = (t+1-t_\text{max}^{(K)})q_\text{max}(t+1) + s_\text{max}^{(K)}$, $q_\text{max}(t+1) = (s_\text{max}^{(K)}-s_\text{max}^{(K-1)})/(t_\text{max}^{(K)}-t_\text{max}^{(K-1)})$, and analogously for $\hat{y}_\text{min}(t+1)$ and $q_\text{min}(t+1)$. Thus, buy(sell)-signals are triggered upon reaching the lower(upper) trading range affinely constructed based upon the last two local minima(maxima).

Our second TR-based controller is referred to as \textsf{TR-Outside}. It triggers trading signals according to:
\begin{equation*}
J(t) = \begin{cases} 1, & \text{if}~\frac{\hat{s}(t+1)-\hat{y}_\text{max}(t+1)}{\hat{y}_\text{max}(t+1)}>\epsilon_\text{TR},\\
0, & \text{if}~\frac{\hat{s}(t+1)-\hat{y}_\text{min}(t+1)}{\hat{y}_\text{min}(t+1)}<-\epsilon_\text{TR},\\
J(t-1), & \text{otherwise}.\end{cases}
\end{equation*} 
Thus, buy(sell)-signals are triggered upon outbraking the upper(lower) trading corridor affinely constructed based upon the last two local maxima(minima).

Note that for the final trading rules of both \textsf{TR-Inside} and \textsf{TR-Outside}, we only employ the last two local maxima and minima, eventhough we derived $s_\text{max}^{(k)}$ in \eqref{eq_def_smax_k} (and similarly $t_\text{max}^{(k)}$, $s_\text{min}^{(k)}$ and $t_\text{min}^{(k)}$) for all $k=1,\dots,K$. This has the reason that a partition according \eqref{eq_def_Twin} can be constructed either with uniform spacings starting at at time $t-T_\text{win}$ and partitioning proceeding forward in time until $t$, \textit{or}, alternatively, starting at time $t$ and partitioning going backward in time. Interestingly, when testing both methods we found the former method to almost always perform better. We attribute this to the residual $\delta$ that typically enlarges the final time-window for $k=K$.

\subsection{Historical optimal-based causal controller}

Given past stock prices historical optimal (HistOpt) trading trajectory can be reconstructed a posteriori with hindsight. This can be done efficiently by graph generation and evaluation. A valid question is whether such optimal trajectories generated up until time $t+1$ with predicted $\hat{s}(t+1)$ as final stock price can also be exploited for real-time (RT) stock trading. Thus, trading signals are
\begin{equation}
J(t) = \begin{cases} \tilde{J}(t), & \text{if}~\tilde{J}(t-\tau)=\tilde{J}(t),~\forall \tau=1,\dots,T_\text{HO}-1,\\
0, & \text{otherwise},\end{cases}
\end{equation} 
where $\tilde{J}(t)$ denotes the historical optimal trading trajectory at time $t$. Tuning parameter $T_\text{HO}$ determines the number of past consecutive identical trading signals necessary to trigger a buy/sell signal. We refer to this controller as \textsf{HistOpt-RT}.

\subsection{Final remark}

All trading controllers discussed within this section were designed to rely on an estimated step-ahead stock price $\hat{s}(t+1)$. Naturally, price data can arbitrarily be shifted by one sampling interval to the past, thereby making the controllers independent of $\hat{s}(t+1)$ while not prohibiting their applicability. The formulation using $\hat{s}(t+1)$ admits for exploitation of any potential good estimate of step-ahead stock prices. Furthermore, it allows a better comparison with the SMPC-based methods, whose \textit{core} is an estimate of one-step ahead stock prices. As will be shown in the subsequent section, this is crucial for both performance and robustness.

\section{Simulation experiments}\label{sec_expts}
\begin{table}[!ht]
\begin{center}
\caption{Overfitting illustrated by means of \textsf{MA-Cross} when optimizing parameters on past data. The parameters are $(p_\text{MA,l},~p_\text{MA,s})$. They are optimized on the training data (November 28, 2014 until November 26, 2015) and then validated for November 27, 2015 until November 25, 2016. The corresponding return performances in percent are denoted by $f_\text{train}$ and $f_\text{val}$. The 30 DAX components are ordered according to increasing absolute performance for the 2-year time period. Buy-and-hold performances $f_\text{val,B\&H}$ are given for comparison. The final row indicates average returns.} 
\label{tab_overfitting}
\bgroup
\def\arraystretch{1.2}
\begin{tabular}{|c|c|c|c|c|}
\hline
\rowcolor[gray]{0.8} Stock & Parameters & $f_\text{train,MA}$ & $f_\text{val,MA}$ & $f_\text{val,B\&H}$ \\\hline
1 & $(1,~95)$ & 0.0 & 61.7 & 7.0 \\
2 & $(20,~74)$ & 2.2 & -29.4 & -25.5\\
3 & $(9,~38)$ & 14.4 & -18.2 & -37.3\\
4 & $(7,~32)$ & 11.4 & -28.1 & -39.5\\
5 & $(21,~37)$ & 35.4 & -19.1 & -5.8\\
6 & $(18,~11)$ & 15.6 & -31.1 & -30.4\\
7 & $(7,~14)$ & 10.5 & -9.3 & -7.9\\
8 & $(20,~29)$ & 54.8 & -20.5 & -20.6 \\
9 & $(9,~18)$ & 24.4 & -30.7 & -35.4\\
10 & $(17,~11)$ & 20.5 & -16.5 & -24.4\\
11 & $(17,~10)$ & 35.9 & -30.5 & -28.7\\
12 & $(14,~26)$ & 13.7 & -9.9 &  4.7 \\
13 & $(18,~25)$ & 41.0 & -4.4 & -8.7\\
14 & $(6,~26)$ & 29.8 & -0.6 & -9.8\\
15 & $(20,~27)$ & 33.9 & -27.2 & -16.2\\
16 & $(4,~21)$ & 26.8 & -30.5 & -12.8\\
17 & $(17,~27)$ & 27.0 & -17.3 & -11.0 \\
18 & $(19,~156)$ & 6.9 & 12.1 & 4.9\\
19 & $(16,~29)$ & 34.1 & 1.8 & 1.7\\
20 & $(12,~39)$ & 11.0 & 1.3 & 8.3\\
21 & $(5,~22)$ & 40.8 & 7.6 & -2.3\\
22 & $(12,~20)$ & 12.7 & -8.4 & 2.5\\
23 & $(11,~21)$ & 27.9 & -24.8 & -10.9\\
24 & $(4,~9)$ & 29.0 & -29.2 & -6.0\\
25 & $(14,~27)$ & 38.2 & 1.6 & 10.5\\
26 & $(17,~10)$ & 21.2 & -12.2 & 2.9 \\
27 & $(9,~23)$ & 32.3 & -29.6 & 7.3\\
28 & $(21,~16)$ & 37.5 & -9.5 & -4.4\\
29 & $(17,~43)$ & 46.5 & -8.4 & 13.1\\
30 & $(13,~19)$ & 48.8 & 15.9 & 50.5 \\\hline
Avg. & -- & 26.1 & -11.4 & -7.5 \\
\hline
\end{tabular}
\egroup
\end{center}
\end{table} 

\begin{table}[!ht]
\begin{center}
\caption{Parameter selections for final simulation experiments.} 
\label{tab_param}
\bgroup
\def\arraystretch{1.2}
\begin{tabular}{|l|l|l|}
\hline
\rowcolor[gray]{0.8} Controller & Parameters & Values \\\hline
 \textsf{QP-E+} & $(M,~\alpha)$ & $(1,~1)$ \\
 \textsf{SMPC-M100} & $(M,~\alpha)$ & $(100,~10)$ \\
 \textsf{SMPC-DH} & $M$ & 100 \\\hline
 \textsf{MA-Cross} & $(p_\text{MA,l},~p_\text{MA,s})$ & $(50,~1)$ \\
 \textsf{MA-Sign} & $(T_\text{MA},~p_\text{MA})$ & $(10,~100)$ \\
 \textsf{TR-Inside} & $(T_\text{win},~p_\text{TR},~\epsilon_\text{TR})$ & $(261,~ 100,~0.01)$ \\
 \textsf{TR-Outside} & $(T_\text{win},~p_\text{TR},~\epsilon_\text{TR})$ & $(261,~20,~0.03)$ \\
 \textsf{HistOpt-RT} & $T_\text{HO}$ & 1 \\
\hline
\end{tabular}
\egroup
\end{center}
\end{table} 

\begin{table}[!ht]
\begin{center}
\caption{Results for the one-year trading period between November 27, 2015 and November 25, 2016. The average total number of trades per year is denoted by $\bar{N}_\text{tr}$. The minimum number of trading days between any two trades is $t_\text{min}$. The average, minimum and maximum performance (all measured in $\%$) are $\bar{f}$, $f_\text{min}$ and $f_\text{max}$. The total percentage of positive returns is $F_\text{pos}$.} 
\label{tab_results}
\bgroup
\def\arraystretch{1.2}
\begin{tabular}{|l|c|c|c|c|c|c|}
\hline
\rowcolor[gray]{0.8} \multicolumn{7}{|l|}{1. Perfect: $\hat{s}(t+1)=s(t+1)$}\\
\hline
 Controller & $\bar{N}_\text{tr}$ & $t_\text{min}$ & $\bar{f}$ & $f_\text{min}$ & $f_\text{max}$ & $F_\text{pos}$ \\\hline
 \textsf{QP-E+} & 56 & 1 & 150.7 & 40.8 & 534.5 & 100 \\
 \textsf{SMPC-M100} & 28 & 1 & 45.7 & 0 & 220.9 & 100 \\
 \textsf{SMPC-DH} & 13 & 1 & 8.2 & 0  & 68.7 & 100 \\\hline
 \textsf{MA-Cross} & 21 & 1 & 27.7 & -3.9 & 95.0 & 93.3 \\
 \textsf{MA-Sign} & 4 & 26 & -8.7 & -32.2 & 50.5 & 23.3 \\
 \textsf{TR-Inside} & 2 & 43 & -2.5 & -27.2 & 15.3 & 60.0 \\
 \textsf{TR-Outside} & 5 & 19 & 7.6 & -31.4 & 55.5 & 70.0 \\\hline
 \textsf{HistOpt-RT} & 75 & 1 & 133.3 & 24.1 & 506.7 & 100 \\
\hline
\rowcolor[gray]{0.8} \multicolumn{7}{|l|}{2. Indifferent: $\hat{s}(t+1)=s(t)$}\\
\hline
 Controller & $\bar{N}_\text{tr}$ & $t_\text{min}$ & $\bar{f}$ & $f_\text{min}$ & $f_\text{max}$ & $F_\text{pos}$ \\\hline
 \textsf{QP-E+} & 0 & 0 & 0 & 0 & 0 & 100 \\
 \textsf{SMPC-M100} & 0 & 0 & 0 & 0 & 0 & 100 \\
 \textsf{SMPC-DH} & 2 & 1 & -4 & -43.6 & 2.0 & 76.7 \\\hline
 \textsf{MA-Cross} & 1 & 10 & 0.7 & -19.7 & 28.9 & 86.7 \\
 \textsf{MA-Sign} & 4 & 26 & -8.7 & -32.2 & 50.5 & 23.3 \\
 \textsf{TR-Inside} & 2 & 42 & -0.5 & -26.5 & 18.9 & 60.0 \\
 \textsf{TR-Outside} & 5 & 21 & -4.2 & -42.5 & 50.8 & 36.7 \\\hline
 \textsf{HistOpt-RT} & 75 & 1 & -52.5 & -73.8 & -30.5 & 0 \\
\hline
\rowcolor[gray]{0.8} \multicolumn{7}{|l|}{3. Random: $\hat{s}(t+1) = s(t) + \eta(t) \frac{1}{t}\sum_{\tau=0}^t |s(\tau)-s(\tau-1)|$}\\
\hline
 Controller & $\bar{N}_\text{tr}$ & $t_\text{min}$ & $\bar{f}$ & $f_\text{min}$ & $f_\text{max}$ & $F_\text{pos}$ \\\hline
 \textsf{QP-E+} & 57 & 1 & -40.1 & -60.7 & 0.4 & 3.3 \\
 \textsf{SMPC-M100} & 20 & 1 & -15.6 & -55.3 & 0.0 & 26.7 \\
 \textsf{SMPC-DH} & 25 & 1 & -22.4 & -54.8 & 0.0 & 23.3 \\\hline
 \textsf{MA-Cross} & 9 & 11 & -5.0 & -36.5 & 68.6 & 23.3 \\
 \textsf{MA-Sign} & 4 & 26 & -8.7 & -32.3 & 50.5 & 23.3 \\
 \textsf{TR-Inside} & 2 & 42 & -0.2 & -26.4 & 26.4 & 56.7 \\
 \textsf{TR-Outside} & 7 & 11 & -6.7 & -39.1 & 29.7 & 30.0 \\\hline
 \textsf{HistOpt-RT} & 116.4 & 1 & -67.2 & -80.3 & -34.7 & 0 \\
\hline
\rowcolor[gray]{0.8} \multicolumn{7}{|c|}{4. Correct Sign: $\hat{s}(t+1) = s(t) + 10\xi(t)\text{sign}(s(t+1)-s(t))$}\\
\hline
 Controller & $\bar{N}_\text{tr}$ & $t_\text{min}$ & $\bar{f}$ & $f_\text{min}$ & $f_\text{max}$ & $F_\text{pos}$ \\\hline
 \textsf{QP-E+} & 117 & 1 & 74.7 & -25.9 & 400.4 & 86.7 \\
 \textsf{SMPC-M100} & 119 & 1 & 57.8 & -29.6 & 338.6 & 90.0 \\
 \textsf{SMPC-DH} & 65 & 1 & 3 & -34.1 & 74.5 & 60.0 \\\hline
 \textsf{MA-Cross} & 21 & 1 & 22.0 & -17.1 & 171.8 & 73.3 \\
 \textsf{MA-Sign} & 4 & 26 & -8.7 & -32.3 & 50.5 & 23.3 \\
 \textsf{TR-Inside} & 6 & 21 & -10.9 & -32.2 & 16.2 & 23.3 \\
 \textsf{TR-Outside} & 51 & 1 & 28.3 & -32.7 & 188.1 & 70.0 \\\hline
 \textsf{HistOpt-RT} & 125 & 1 & 69.3 & -28.9 & 380.7 & 83.3 \\
\hline
\rowcolor[gray]{0.8} \multicolumn{7}{|l|}{5. Wrong Sign: $\hat{s}(t+1) = s(t) - 10\xi(t) \text{sign}(s(t+1)-s(t))$}\\
\hline
 Controller & $\bar{N}_\text{tr}$ & $t_\text{min}$ & $\bar{f}$ & $f_\text{min}$ & $f_\text{max}$ & $F_\text{pos}$ \\\hline
 \textsf{QP-E+} & 117.4 & 1 & -93.5 & -98.7 & -87.6 & 0 \\
 \textsf{SMPC-M100} & 119 & 1 & -93.4 & -98.6 & -88.5 & 0 \\
 \textsf{SMPC-DH} & 120 & 1 & -93.7 & -98.7 & -88.8 & 0 \\\hline
 \textsf{MA-Cross} & 11 & 3 & -16.9 & -40.8 & 2.7 & 3.3 \\
 \textsf{MA-Sign} & 4 & 26 & -8.7 & -32.3 & 50.5 & 23.3 \\
 \textsf{TR-Inside} & 5 & 28 & -3.1 & -28.7 & 26.1 & 40.0 \\
 \textsf{TR-Outside} & 56 & 1 & -64.3 & -97.1 & -19.3 & 0 \\\hline
 \textsf{HistOpt-RT} & 136 & 1 & -94.9 & -98.7 & -91.4 & 0 \\  
\hline
\rowcolor[gray]{0.8} \multicolumn{7}{|l|}{Global Optimum (Trading w/ Hindsight)/Buy-and-Hold}\\
\hline
 Controller & $\bar{N}_\text{tr}$ & $t_\text{min}$ & $\bar{f}$ & $f_\text{min}$ & $f_\text{max}$ & $F_\text{pos}$ \\\hline
\textsf{HistOpt}  & 42 & 1 & 192.6 & 64.7 & 609.5 & 100 \\
\textsf{Buy-and-Hold}  & 1 & 0 & -7.5 & -39.5 & 50.5 & 36.7 \\
\hline
\end{tabular}
\egroup
\end{center}
\end{table}

For simulation experiments, we employ the stock prices of the 30 components of the German stock market index DAX between November 28, 2014 and November 25, 2016. For closed-loop trading we only considered the past year and initialized $t=0$ on November 27, 2015. Nevertheless, previous price data was still relevant for the generation of measures such as moving averages at $t=0$. All data was drawn from \texttt{finance.yahoo.com}, see Figure \ref{fig:fig1} for visualization. Throughout, proportional transaction costs of $1\%$ are assumed. All simulations were run on a laptop running Ubuntu 14.04 equipped with an Intel Core i7 CPU @2.80GHz$\times$8, 15.6GB of memory and using Python 2.7.

\subsection{Closed-loop trading results}

Each stock is traded separately and the portfolio with transition dynamics according to Section \ref{sec_transitionDynamics} is initialized with $Z(0)=\begin{bmatrix} 0 & M_0 & 0 & M_0\end{bmatrix}$ where $M_0=100000$\EUR. We compare eight different controllers and five different methods that we use for the prediction of $\hat{s}(t+1)$. In addition we state the results for a \textit{buy-and-hold} strategy (investing maximally into the stock at $t=0$ and consequently holding the investments throughout), and for the global optimal trading result (trading with hindsight). We assume proportional transaction costs $\epsilon=0.01$ identical for both buying and selling of stocks. Performance is defined by $f=\frac{W(T)-M_0}{M_0}100$ with $T$ the final trading date.   

Regarding parameter selections for the genetic algorithms, we tested three settings. First, we optimized parameters on training data (November 28, 2014 until November 26, 2015) and then validated for November 27, 2015 until November 25, 2016. Thus, for each stock and for each controller individual parameters were selected . Second, we recursively updated parameters. Thus, every 100 days (we also tested 20 and 50 days) we recomputed parameters optimized on past data of one year at that time. Third, we arbitrarily chose a fixed parameter set for each controller and used this for the trading of \textit{all} 30 stocks. For both the first and the second approach strong overfitting could be observed, see Table \ref{tab_overfitting} for illustration. We therefore opted for fixed parameter selections for the trading of all stocks. More conservative parameter selections, such as, e.g., larger MA-windows $p_\text{MA,l}$, performend on average better. For \textsf{HistOpt-RT} we intentionally chose $T_\text{HO}=1$, which is the most aggressive but least robust choice as outlined in the following. The parameter selections employed for final simulation experiments are summarized in Table \ref{tab_param}. 

Closed-loop trading results are summarized in Table \ref{tab_results}. Because of the importance of step-ahead predictions and for robustness considerations, we compare five different versions for $\hat{s}(t+1)$. Ideally (but unrealistically in general), $s(t+1)$ is estimated perfectly, i.e., $\hat{s}(t+1)=s(t+1)$. This case is noncausal. Nevertheless, it serves as an important benchmark. Case 4 and 5 (``correct'' and ``wrong sign prediction'') are likewise noncausal since $s(t+1)$ is not known at time $t$. Guiding notion for their introduction was to analyze influence of correct trend prediction (up or down) for the price one time-step ahead but without knowledge of exact level of price rise or fall. We therefore add multiplicative time-varying perturbation $10\xi(t)$ with $\xi(t)\sim\mathcal{U}(0,1)$ uniformly distributed. Case 2 (``indifferent'') uses the current stock price as the estimate for the ext time-step ahead. Case 3 (``random'') randomly perturbs $s(t)$ as the estimate for $\hat{s}(t+1)$, whereby $\eta(t)\sim\mathcal{N}(0,1)$ normally distributed. Results are discussed in the next section. Importantly, we remark that only 36.7\% of the 30 DAX components rose, i.e., $s(T)>s(0)$, for the one-year trading period considered between November 27, 2015 and November 25, 2016.

\subsection{Discussion}

Several observations can be made from Table \ref{tab_results}. Let us first discuss the results for SMPC-based stock trading and \textsf{HistOpt-RT}. For the ideal case of perfect $s(t+1)$ knowledge truely excellent results can be obtained. Both for \textsf{QP-E+} and \textsf{HistOpt-RT}. Performances are even in range with the global optimum (\textsf{HistOpt}) despite only \textit{one-step} ahead price knowledge. By reduction of $\alpha$ from 10 to 1, quasi identical performance to \textsf{QP-E+} is achieved by \textsf{SMPC-M100}, i.e., $(\bar{f},f_\text{min},f_\text{max})=(139.4,42.7,533.7)$. We selected $\alpha=10$ to illustrate its role in adding robustness. This becomes apparent for the random $\hat{s}(t+1)$ prediction method (case 3): while $\bar{f}=-40.1\%$ for \textsf{QP-E+}, it is $\bar{f}=-15.6\%$ for \textsf{SMPC-M100}, and additionally yielding 26.7\% positive returns overall. For the indifferent prediction $\hat{s}(t+1)=s(t)$, \textsf{QP-E+} and \textsf{SMPC-M100} \textit{never} enter a trade. This was expected. Both optimization problem formulations essentially rely on the mean difference between $s(t)$ and $\hat{s}(t+1)$. Characteristically they do not consider \textit{any} past data points except the current price $s(t)$. This is in contrast to \textsf{HistOpt-RT} where all available past data up until $t$ is searched for the optimal traing trajectory. \textsf{SMPC-DH} was found to not be competitive with the other two SMPC-based controllers (\textsf{QP-E+} and \textsf{SMPC-M100}), neither with respect to performance nor robustness. Nevertheless, its framework is favorable in that more (and better) heuristics can easily be incorporated by adjusting reference scenarios $p^j(t+1),~j=1,\dots,M$. 

Of great relevance to SMPC-based trading strategies and \textsf{HistOpt-RT} are the experiments for correct and wrong sign predictions (case 4 and 5). Importantly, they indicate that perfect one-step ahead \textit{sign} predictions of price changes $s(t+1)-s(t)$ are sufficient for excellent results. Thus, the precise level of increase or decrease in stock prices is not necessarily required. For illustration, consider the average gain of 74.7\% \textit{per stock} for \textsf{QP-E+} despite the fact that only 36.7\% of all 30 DAX components actually rose during the past year and moreover on average yielding minus 7.5\% (see the Buy-and-Hold strategy). Even more important are the results for case 5, i.e., when at every trading instant wrongly predicting the direction of change in stock prices. For \textit{all} SMPC-based trading methods and \textit{HistOpt-RT}, \textit{after just one year}, \textit{at least} 93.5\% of all initial wealth is lost. 

Finally, note that for the SMPC-based methods and \textsf{HistOpt-RT} the minimum time-span between any two trades is always 1 day (except for case 2 when there are no trades at all). Furthermore, the average number of trades per year, $\bar{N}_\text{tr}$, is considerably larger in comparison to the genetic algorithms. 

Throughout experiments, more robust performance could be observed for the genetic algorithms. For the given parameter choices, \textsf{MA-Cross} appeared to be best suited to exploit potential knowledge of future stock prices (case 1). Encouraging are the returns for \textsf{MA-Cross} and \textsf{TR-Inside} for the causal prediction case, i.e., $\hat{s}(t+1)=s(t)$. Despite the fact that only 36.7\% of all 30 DAX components were actually rising since last year, the two controllers yielded positive returns for 86.7\% and 60\%, respectively. For random price-ahead predictions (case 3), the performance of all four MA- and TR-based controllers was comparable to the \textsf{Buy-and-Hold} method; with \textsf{TR-Inside} performing best on average and with respect to worst-case losses.

Before concluding, let us remark some realistic success ratios reported in the literature for correct sign predictions of step-ahead price difference $s(t+1)-s(t)$. In \cite{kim2003financial}, support vector machines (SVM) in combination with 12 technical indicators (such as Williams \%R, stochastic \%K, disparity, etc.) are used to predict the direction of change in the daily Korea composite stock price index (KOSPI). For validation data and their best tuning parameter choices, they report a prediction performance between 50.1\% and 57.8\%. The same author mentioned similar results in earlier work \cite{kim2000genetic}.

\section{Conclusion}\label{sec_conclusion}

For stock trading, the general class of genetic algorithms appears more suitable than methods based on stochastic model predictive control. The former class is signficantly more robust. A SMPC-approach is justifiable only for consistently perfect prediction of direction of price changes. This is not achievable in practice. The relations and differences between using SMPC for dynamic hedging and stock trading were discussed.

Findings motivate the following:
\begin{enumerate}
\item A detailed analysis of scenarios when MA- and TR-based algorithms fail and succeed, respectively.
\item An artificial and automated generation of genetic trading algorithms to further improve performance and robustness \cite{allen1999using}.
\item The usage of options for their predictable worst-case loss~\cite{plessen2017parallel}.
\end{enumerate}
Subject of future research are the application of genetic trading algorithms to both multi-asset portfolio optimization and dynamic option hedging.

\nocite{*}
\bibliographystyle{ieeetr}
\bibliography{myref}

\end{document}